\documentclass[galaxies,review,accept,pdftex,oneauthor]{Definitions/mdpi} 
\firstpage{1} 
\pubvolume{13}
\issuenum{2}
\articlenumber{33}
\pubyear{2025}
\copyrightyear{2025}
\externaleditor{Dominic Bowman}
\datereceived{14 February 2025 } 
\daterevised{26 March 2025 } 
\dateaccepted{28 March 2025 } 
\datepublished{2 April 2025} 
\hreflink{https://doi.org/ 10.3390/galaxies13020033} 

\Title{Red Supergiants as Supernova Progenitors}

\TitleCitation{Red Supergiants as Supernova Progenitors}

\Author{{Schuyler D.~Van Dyk}
 \orcidA{}}

\AuthorNames{Schuyler D.~Van Dyk}

\AuthorCitation{Van Dyk, S.D.}

\address[1]{%
{Caltech/IPAC}, {Mailcode} 
 {100-22,} 
 Pasadena, CA 91125, USA; vandyk@ipac.caltech.edu; Tel.: +1-626-395-1881}

\abstract{The inevitable fate of massive stars in the initial mass range of $\approx$8--$30\ M_{\odot}$ in the red supergiant (RSG) phase is a core-collapse supernova (SN) explosion, although some stars may collapse directly to a black hole. We know that this is the case, since RSGs have been directly identified and characterized for a number of supernovae (SNe) in pre-explosion archival optical and infrared images. RSGs likely all have some amount of circumstellar matter (CSM), through nominal mass loss, although evidence exists that some RSGs must experience enhanced mass loss during their lifetimes. The SNe from RSGs are hydrogen-rich Type II-Plateau (II-P), and SNe II-P at the low end of the luminosity range tend to arise from low-luminosity RSGs. The typical spectral energy distribution (SED) for such RSGs can generally be fit with a cool photospheric model, whereas the more luminous RSG progenitors of more luminous SNe II-P tend to require a greater quantity of dust in their CSM to account for their SEDs. The SN II-P progenitor luminosity range is 
$\log(L_{\rm bol}/L_{\odot})\sim4.0$--5.2. The fact RSGs are known up to $\log(L_{\rm bol}/L_{\odot}) \sim 5.7$ leads to the so-called ``RSG problem'', which may, in the end, be a result of small number of available statistics to date.}


\begin{document}

\section{Introduction}\label{sec:intro}

It had long been speculated that red supergiants (RSGs) were the progenitors of supernovae (SNe), catastrophic explosions arising from the gravitational-collapse endpoints of stars with initial, zero-age main sequence masses $M_{\rm ini} \gtrsim 8\ M_{\odot}$ \cite{Woosley1986} (see Ekstr{\"o}m \& Georgy in this volume). As~such, the resulting SNe should be dominated by strong hydrogen lines in their spectra~\cite{Branch1981}, given the massive RSG hydrogen envelope, i.e.,~they are SNe of Type II~\cite{Minkowski1941}, and~the light curves should exhibit an extended ``plateau'' in their luminosity~\cite{Falk1973,Chevalier1976,Arnett1980}. More modern theoretical analyses have further strengthened this connection, through detailed modeling of both the SN photometric and spectroscopic evolution, e.g.,~\cite{Dessart2008,Hillier2012,Dessart2013,Pejcha2015,Hillier2019,Dessart2019,Goldberg2019}.

One must consider the observed properties of SNe in the context of their progenitors. They are inextricably linked. Substantial heterogeneity exists in both the overall shapes and the peak luminosities of Type II SNe (SNe~II) \cite{Anderson2014}, with~implications for diversity in progenitor properties at the time of explosion. Classically, SN~II light curves were separated into II-Plateau (II-P) and II-Linear (II-L) \cite{Barbon1979}, with~possible spectroscopic differences exemplified by a weaker P-Cygni profile in the H$\alpha$ line~\cite{Schlegel1996}. More recently, the~light-curve shape distinction has become less clear with significantly more data available~\cite{Anderson2014,Faran2014,Valenti2016}, although~large samples appear to indicate that SNe with higher expansion velocities have higher luminosities, higher radioactive $^{56}$Ni masses, shorter ($\lesssim$100-day) plateau durations, and~more rapidly declining light curves~\cite{Gutierrez2017,Hiramatsu2021a}. At~the other extreme are low-luminosity SNe~II, with~peak luminosities at least an order of magnitude lower than average, underluminous exponential declines in the light-curve tail, $\sim$10\% of the average $^{56}$Ni mass, and~significantly lower expansion velocities~\cite{Zampieri2003,Pastorello2004}.
 
{Recent sophisticated modeling has impressively simulated SN II-P light curves through radiative transport of shock-deposited and radioactively powered energy through the RSG stellar ejecta~\cite{Tsang2020}, successfully reproducing the observations of the radiative breakout of the shock wave through the outer RSG envelope~\cite{Goldberg2022}.}

Neutrino heating, together with neutrino-driven---albeit chaotic--- turbulence, appears to be the main mechanism driving explosion in massive stars, although~an understanding of stellar evolution to core collapse has not yet converged~\cite{Burrows2024}.
{Whereas} empirical (and possibly theoretical) correlations may exist between initial progenitor mass; ejecta expansion velocity; and, therefore, explosion energy~\cite{Nomoto2006,Poznanski2013,Muller2016,Martinez2019,Burrows2020} (however, see~\cite{Pejcha2015a,Pejcha2015b}), recent 3D core-collapse SN models point more toward explosion energy as a function of, say, progenitor mantle binding energy and, more importantly, a~core \mbox{``compactness parameter'' \cite{OConnor2011}} rather than initial progenitor mass~\cite{Burrows2024}. {We point out that explosions have also been modeled as jet-driven~\cite{Couch2009,Papish2015,Soker2022}.}

We know that many RSGs are often-irregular, long-period variables (LPVs; \mbox{e.g.,~\cite{Stothers1971,Kiss2006,Soraisam2018})}. We also know that many, if~not all, RSGs have at least some circumstellar matter (CSM, e.g.,~\cite{VanLoon1999}) beyond their photospheres, often dusty~\cite{Massey2005,Verhoelst2009}, as~a result of mass loss during this phase of stellar evolution~\cite{Mauron2011,Humphreys2020}. 
We observe evidence for a CSM layer above the photosphere via short-lived so-called ``flash'' emission features in early-time optical spectra for a number of SNe II~\cite{Khazov2016}. The~interaction of the SN shock with the dense CSM can lead to enhanced luminosity in the early-time light curves~\cite{Moriya2017,Morozova2020}. In~fact, if~the CSM is sufficiently dense, the~breakout of the SN shock can be significantly delayed~\cite{Forster2018,Zimmerman2024}.

A question then emerges: What is the origin of the CSM around RSGs, and, then, at what rate and at what interval does it emerge before explosion, especially when the CSM is particularly dense? Theoretical evolutionary models include baseline empirical prescriptions of mass loss~\cite{deJager1988,Nieuwenhuijzen1990} ($\sim$$10^{-6}\ M_{\odot}$ yr$^{-1}$; different, higher mass-loss rate prescriptions have been presented elsewhere~\cite{vanLoon2005}; see also van Loon in this volume and~\cite{Beasor2020}). However, these rates are not nearly high enough to account for the amount and densities of the CSM that are inferred from a number of SN observations, which, in several, cases also imply short-time-scale (days to decades) mass-loss episodes prior to~explosion.

Enhanced mass loss ($\sim$$10^{-4}$--$10^{-2}\ M_{\odot}$ yr$^{-1}$) can be driven pulsationally by internal convective shocks~\cite{Josselin2007} induced by nuclear flashes~\cite{Moriya2011} or convectively driven wave heating~\cite{Shiode2014,Fuller2017}, which, toward the end of the RSG phase during late-stage nuclear burning, could lead to higher mass-loss rates~\cite{Beasor2016} or~eruptive outbursts during the final year or years prior to explosion~\cite{Yaron2017,Bruch2021,Davies2022}, when turbulence and pulsations are most vigorous. (A ``superwind'' could develop from internal dynamical instabilities; however, this may only be the case for the most massive RSGs~\cite{Yoon2010}.) Such a precursor event has been observed for at least one normal SN II-P~\cite{JacobsonGalan2022}. Other recent studies have provided alternatives to outbursts, e.g.,~a late-time effervescent {zone} or a wave-driven chromospheric layer~\cite{Fuller2024} to account for the immediate dense CSM above RSGs. Binary interaction may offer yet another mechanism~\cite{Matsuoka2024}.

Several indirect observational constraints on SN progenitors include the modeling of SN light curves via progenitor population synthesis~\cite{Eldridge2018}, probing late-time ejecta~\cite{Milisavljevic2012,Jerkstrand2012}, analyzing circumstellar gas ionized by the early ultraviolet/X-ray radiation from the SN \mbox{blast~\cite{Gezari2008,GalYam2014,Khazov2016},} and~inferring ages and turnoff masses from the SN's local stellar and interstellar environment (\cite{Anderson2012,Maund2017,Kuncarayakti2018,Williams2018}; however, see~\cite{Zapartas2021}). Of course, the~surest means to understand the nature of the progenitor is to directly identify and characterize it as~it was some time before it exploded. The~first SN II to have its progenitor identified in this way was SN 1987A, although~the SN was peculiar in its properties. In~addition, ironically, the~progenitor was identified as {\it blue}, not red, supergiant Sanduleak $-69^{\circ}$ 202~\cite{Sonneborn1987,Walborn1989} (the unanticipatedly unusual progenitor could possibly be explained by a merger of two stars~\cite{Podsiadlowski1990}). The~first SN II-P progenitor to be characterized from pre-SN imaging as an RSG was SN 2003gd~\cite{VanDyk2003a,Smartt2004}. Subsequent identifications up to a certain point in time have been reviewed and discussed elsewhere~\cite{Smartt2009a,Smartt2009b,Smartt2015,VanDyk2017}. In~the remainder of this article, we provide a further and more current~review.


\section{Direct Identification of RSGs as SN~Progenitors}\label{sec:direct}

A summary of all of the direct identifications of SN II-P progenitors to date is given in Table~\ref{tab1}. Like SN 1987A, the~host galaxies of SN 2004et~\cite{Li2005,Crockett2011}, SN 2008bk~\cite{Mattila2008,VanDyk2012a,ONeill2021}, and~SN 2012A~\cite{Prieto2012,Tomasella2013} were nearby enough ($\lesssim$10 Mpc) that the progenitors could be identified in ground-based image data (this was strikingly true for SN 2008bk, whose host is at 3.4 Mpc, and~data of high quality in multiple photometric bands were available). All of the remaining examples involve images taken prior to each SN with the high spatial resolution of the {\sl Hubble Space Telescope\/} ({\sl HST}). The~SN 2003gd progenitor identification was a hybrid of {\sl HST\/} data at shorter optical wavelengths and ground-based images at longer wavelengths. A~number of cases also included infrared data from the {{\sl Spitzer Space Telescope}}.

\begin{table}[H] 
\caption{Direct identification of RSGs as SN~progenitors.\label{tab1}}
\begin{tabularx}{\textwidth}{CCCCC}
\toprule
\textbf{SN}	& \textbf{Host}	& \boldmath{$\log(L_{\rm bol}/L_{\odot})$} & \boldmath{{$T_{\rm eff}$ \textbf{(K)}}} & \textbf{Refs.}\\
\midrule
2003gd	& NGC 628	      & 4.3                     & $\sim$3500 & \cite{VanDyk2003a,Smartt2004}\\
2004A   & NGC 6207        & $\sim$4.2--4.7          & $\sim$2800--5900        & \cite{Hendry2006}\\
2004et  & NGC 6946        & 4.54                    & $\sim$4300--6150        & \cite{Li2005,Crockett2011}\\ 
2005cs  & NGC 5194        & $\sim$4--4.5            & $\sim$2950--4500        & \cite{Maund2005a,Li2006}\\
2006my  & NGC 4651        & $\sim$4.2--4.8, $<$$5.1$  & $\lesssim$3750        & \cite{Li2007,Leonard2008,Crockett2011}\\
2008bk  & NGC 7793        & 4.5--4.6                & $\sim$3500--3700        & \cite{Mattila2008,VanDyk2012a,ONeill2021}\\
2008cn  & NGC 4603        & $\lesssim$5             & $\gtrsim$3160        & \cite{EliasRosa2009,Maund2015}\\
2009hd \textsuperscript{1} & NGC 3627  & $\lesssim$5.05 & $\lesssim$5200      & \cite{EliasRosa2011}\\
2009ib  & NGC 1559        & 5.1                     & $\sim$3400        & \cite{Takats2015}\\
2012A   & NGC 3239        & 4.7                     & $\sim$4250--3400        & \cite{Prieto2012,Tomasella2013}\\
2012aw  & NGC 3351        & 4.8--5.0                & $\sim$3400--3900        & \cite{VanDyk2012b,Fraser2012,Kochanek2012}\\
2012ec  & NGC 1084        & 5.15                    & $<$4000        & \cite{Maund2013}\\
2013ej \textsuperscript{1} & NGC 628  & 4.46--4.85   & $\sim$3400--4000        & \cite{Fraser2014}\\
2017eaw & NGC 6946        & 4.9, 5.07               & $\sim$2500--3300        & \cite{Kilpatrick2018,VanDyk2019}\\
2018zd \textsuperscript{2} & NGC 2146 &$\sim$4.5--5.1 & $\sim$3500 \textsuperscript{3}       & \cite{Hiramatsu2021b}\\ 
2018aoq & NGC 4151        & 4.7                     & $\sim$3500        &\cite{ONeill2019}\\
2020jfo & NGC 4303        & 4.1 ? \textsuperscript{4}                    & $\sim$2900        &\cite{Sollerman2021,Kilpatrick2023a}\\
2022acko & NGC 1300       & 4.3--4.5                & $\sim$3500--3565        &\cite{VanDyk2023b}\\
2023ixf & NGC 5457        & 4.7--5.5                & $\sim$2770--4000        & \cite{Kilpatrick2023b,Jencson2023,Pledger2023,Niu2023,Soraisam2023,Neustadt2024,Xiang2024a,Ransome2024,VanDyk2024,Qin2024}\\ 
2024ggi & NGC 3621        & 4.9                     & $\sim$3290        & \cite{Xiang2024b}\\
{2024abfl} & NGC 2146 & $\sim$4.6 ? \textsuperscript{4}  &  $\sim$3600 ? \textsuperscript{4} & \cite{Luo2024}\\
\bottomrule
\end{tabularx}
\noindent{\footnotesize{\textsuperscript{1} Possible II-L. \textsuperscript{2} Peculiar II-P, possible SAGB progenitor. \textsuperscript{3} If the progenitor is similar to SAGB candidate MSX SMC 055, possibly $T_{\rm eff} \sim 2500$~K~\cite{Groenewegen2009}. \textsuperscript{4} Question mark, ``?'', denotes uncertainty and necessity to confirm further.}}
\end{table}

The precise location of the progenitor is generally established based on high-resolution images of the SN itself, usually when the SN is still bright, using either the {\sl HST\/} or adaptive optics-assisted ground-based observations (AO). A~notable exception is SN 2022acko, for~which the SN was serendipitously imaged by the {\sl James Webb Space Telescope\/} ({\sl JWST}); those data were astrometrically aligned with the pre-SN {\sl HST\/} imaging to pinpoint the progenitor site~\cite{VanDyk2023b}.

\begin{figure}[H]
\begin{adjustwidth}{-\extralength}{0cm}
\centering
\includegraphics[width=15.5cm]{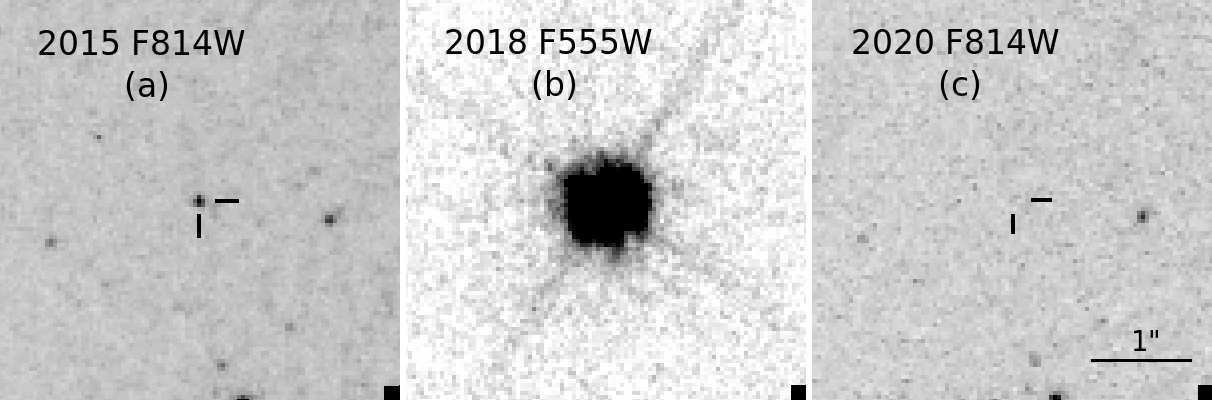}
\end{adjustwidth}
\caption{(\textbf{a}) Portion of an {\sl HST\/} image of the host galaxy of SN 2018aoq obtained in the F814W band in December 2015. (\textbf{b}) Portion of an {\sl HST\/} image of the SN itself obtained in the F555W band in April 2018. (\textbf{c}) Portion of an {\sl HST\/} image of the SN field obtained in the F814W band in December 2020. The~position of the SN site in panels (\textbf{a},\textbf{c}) is indicated with tick marks. The~image data were all obtained with the Wide-Field Camera 3 (WFC3) instrument. The~panels are all shown at the same scale and orientation. North is up, and~east is to the left. (See also~\cite{ONeill2019,VanDyk2023a}).\label{sn18aoq}}
\end{figure}

In Figure~\ref{sn18aoq}, we show one example of direct progenitor identification for~SN 2018aoq in NGC 4151 (see also~\cite{ONeill2019}). The leftmost panel shows the RSG progenitor in the {\sl HST\/} F814W ($\sim$$I$) band, the~site of which was serendipitously imaged more than 2 years prior to its explosion. The~precise location of the progenitor was pinpointed using an image of the SN itself we obtained with the {\sl HST\/} in 2018 in the F555W band ($\sim$$V$). We revisited the site again in 2020 with the {\sl HST} in the F814W band, confirming that the RSG was, indeed, the progenitor based on its disappearance~\cite{VanDyk2023a}.

In Table~\ref{tab1}, we provide both the host galaxy name and the inferred bolometric luminosity of~the RSG progenitor relative to the Sun ($\log[L_{\rm bol}/L_{\odot}]$). In~the literature on each of the progenitors, the~initial mass($M_{\rm ini}$), is generally inferred for the stars. However, we resist including these here and list just the luminosity instead. This is since the mapping from luminosity to initial mass, via theoretical stellar evolutionary tracks, can be fraught with additional uncertainties and open to interpretation~\cite{Davies2020a}, particularly given that the various available model tracks~\cite{Bressan2012,Choi2016,Stanway2018} for a given $M_{\rm ini}$ do not all terminate at the same locus on the Hertzsprung--Russell diagram, due, at least in part, to different input assumptions made in the~modeling. 

Many of the cases involve detection of the progenitor only in one band, with~upper limits on detection in other bands, especially early on in the pursuit of direct progenitor identifications, e.g.,~SN 2005cs~\cite{Maund2005a,Li2006}. Typically, these one- or two-band detections have since been converted to luminosity estimates via bolometric corrections for \mbox{RSGs~\cite{Levesque2005,Levesque2006,Davies2018}}. When more than two photometric bands are available, model fits can be applied to the resulting observed spectral energy distributions (SEDs), then integrated over wavelengths to obtain luminosity~estimates.

In practice, a limit exists on how far away the host can be while still being able to detect or resolve an individual luminous star. For the {\sl HST\/}, that limit is $\lesssim$20--30 Mpc, beyond~which crowding and confusion become insurmountable issues; for ground-based optical and near-infrared (and~even the {\sl Spitzer\/} telescope in the mid-infrared range), that limit is far more restrictive. In~nearly equal numbers to the progenitor detections are non-detections, many as a result of the depth (or lack thereof) of the pre-SN imaging, given either the exposure time, the~host distance, or~both, as listed in Table~\ref{tab2}. In~a few of these cases, the resulting upper limits provided useful constraints on the progenitor's luminosity, which are provided in the~table.

\begin{table}[H] 
\caption{Upper limits on progenitor~detection.\label{tab2}}
\begin{tabularx}{\textwidth}{CCCC}
\toprule
\textbf{SN}	& \textbf{Host}	& \boldmath{$\log(L_{\rm bol}/L_{\odot})$} & \textbf{Refs.}\\
\midrule
1999an                     & IC 755   & \ldots                           & \cite{VanDyk2003b,Maund2005b}\\
1999br                     & NGC 4900 & \ldots                           & \cite{VanDyk2003b,Maund2005b}\\
1999em                     & NGC 1637 & \ldots                           & \cite{Smartt2002}\\
1999ga \textsuperscript{1}  & NGC 2442 & \ldots                           & \cite{Pastorello2009b}\\
1999gi                     & NGC 3184 & $\lesssim$$5.1$                & \cite{Smartt2001}\\
2001du                     & NGC 1365 & \ldots                           & \cite{VanDyk2003c,Smartt2003,Maund2005b}\\
2002hh                     & NGC 6946 & \ldots                           & \cite{Smartt2009a}\\
2003ie \textsuperscript{2}  & NGC 4051 & \ldots                           & \cite{Smartt2009a}\\
2004am \textsuperscript{3}  & NGC 3034 & \ldots                           & \cite{Mattila2013}\\
2004dg                     & NGC 5806 & $\gtrsim$$4.5$, $\lesssim$$5.1$ & \cite{Smartt2009a}\\
2004dj \textsuperscript{3}  & NGC 2403 & \ldots                           & \cite{MaizApellaniz2004}\\
2006bc                     & NGC 2397 & $\gtrsim$$4.8$, $\lesssim$$5.1$ & \cite{Smartt2009a}\\
2007aa                     & NGC 4030 & $\gtrsim$$4.5$, $\lesssim$$5.1$ & \cite{Smartt2009a}\\
2009H                      & NGC 1084 & \ldots                           & this~work\\
2009N                      & NGC 4487 & $\gtrsim$$4.5$, $\lesssim$$5.1$ & this~work\\
2016cok                    & NGC 3627 & \ldots                           & \cite{Kochanek2017}\\
2018ivc \textsuperscript{4} & NGC 1068 & $\gtrsim$$4.5$, $\lesssim$$5.1$ & \cite{Bostroem2020}\\
2019mhm                    & NGC 6753 & \ldots                           & \cite{Vazquez2023}\\
2020fqv                    & NGC 4568 & \ldots                           & \cite{Tinyanont2022}\\
2020pvb \textsuperscript{5} & NGC 6993 & \ldots                           & \cite{EliasRosa2024}\\
2021yja                    & NGC 1325 & \ldots                           & \cite{Vasylyev2022}\\ 
\bottomrule
\end{tabularx}
\noindent{\footnotesize{\textsuperscript{1} Possible II-L. \textsuperscript{2} Peculiar II-P{?}. \textsuperscript{3} In stellar cluster. \textsuperscript{4} Possible II-L or IIb. \textsuperscript{5} Possible ``IIn-P''.}}
\end{table}

\begin{figure}[H]
\includegraphics[width=10.5 cm]{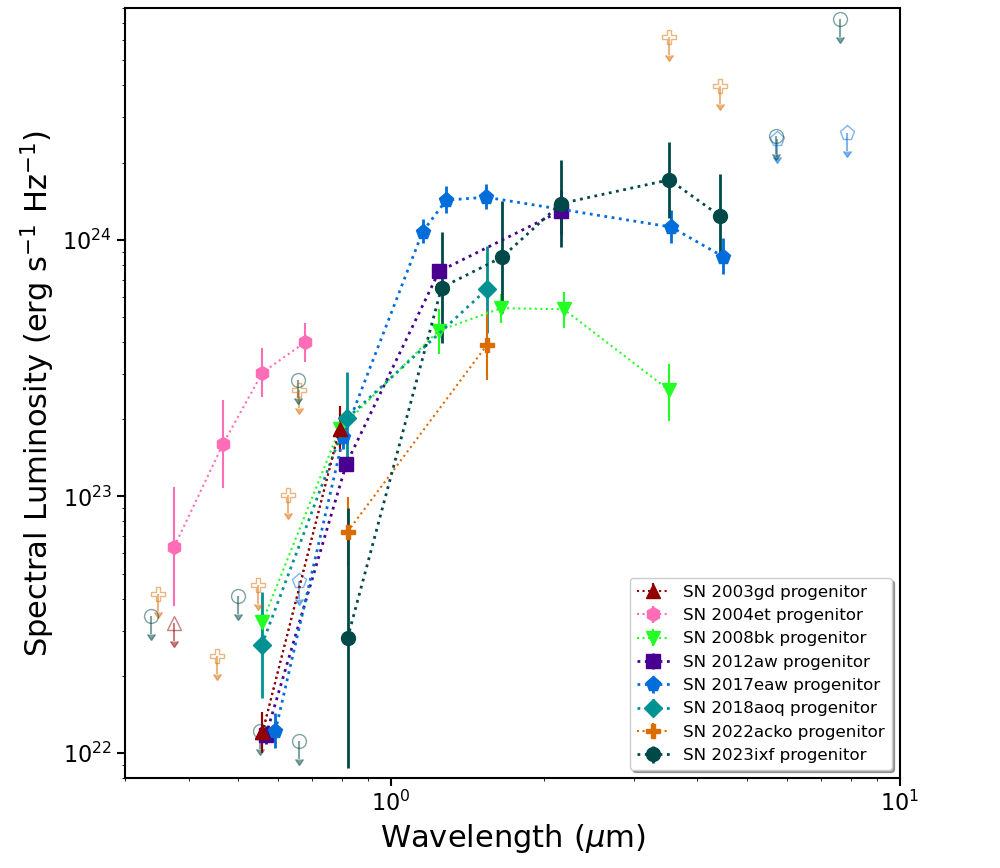}
\caption{A comparison of the spectral energy distributions (SEDs) for several resolved RSG progenitors of SNe II-P (see Table~\ref{tab1}). {Photometric bands for which there are measurements are shown as filled symbols, and upper-limit estimates are presented as~open symbols.} SEDs are all corrected for both galactic foreground and internal host galaxy reddening, as well as the assumed distances to the~hosts.\label{sedcomp}}
\end{figure}

We show absolute SEDs for several of the detected progenitors listed in Table~\ref{tab1} in Figure~\ref{sedcomp}. The~observed SEDs were all corrected both for assumptions of galactic foreground and internal host galaxy reddening, as~well as for the assumed distance to each of the hosts. As~one can see, first, the SEDs were not all been equally sampled across wavelengths, since we are completely beholden to whatever available archival imaging data serendipitously included the SN site. Secondly, with~the caveat that the statistics are still quite small, the~SEDs have generally different shapes relative to one another; the variety of shapes most likely arise from differing amounts of intrinsic reddening in (i.e., dustiness of) the circumstellar environments of the progenitors. This is not entirely surprising, since we do not expect the apparent SEDs of RSGs to be the same, given the rich variety of SED shapes for, e.g.,~galactic RSGs~\cite{Verhoelst2009}. We should not expect all massive stars to have the same properties at the end of their lives; therefore, each RSG is an individual. Lastly, the~detected progenitors span a range of at least an order of magnitude in luminosity, as~can be seen as well in Table~\ref{tab1}, from~$\log(L_{\rm bol}/L_{\odot}) \sim 4$ to 5.5. This is, to~the first order, consistent with the expected range of $M_{\rm ini}$ ($\sim$$8$--$30\ M_{\odot}$) for massive stars that terminate as RSGs (however, we discuss this further in Section~\ref{sec:rsg_problem}). 

We must stress here that we have been using the term ``progenitor'' rather cavalierly so far. A star identified at the precise location of an SN is only a {\em candidate\/} progenitor until~the star has demonstrably vanished. This requires patient vigilance to determine whether the SN has faded sufficiently below the luminosity of the candidate, with~the wait time occasionally exacerbated by any lingering light as a result of CSM interaction. For~many of the SNe in Table~\ref{tab1}, the candidates have been shown to have disappeared, reinforcing the likelihood that these were the actual \mbox{progenitors~\cite{Maund2009,VanDyk2013,Maund2014,Maund2015,VanDyk2015,Fraser2016,VanDyk2023a}.} One important caveat to this is that we cannot outright eliminate the possibility that dust has formed around the SN and obscured or~otherwise dimmed the~light. We expect dust to form over the time span of years in the SN ejecta from massive star explosions~\cite{Sarangi2015}. Additionally, unresolved binaries or multiple-star systems certainly can appear as single stars at the distances of the host galaxies. We would expect this to occur quite frequently, since the multiplicity fraction is $\gtrsim$$70$\% for (primary) stars with $M_{\rm ini} \gtrsim 8\  M_{\odot}$ \cite{Chen2024}. It is also possible that an identified progenitor is a chance superposition of two or more stars, again, given the host distances. In~fact, the~yellow star associated with SN 2008cn~\cite{EliasRosa2009} appears, in retrospect, to have been a combination of a blue and a red (probably supergiant) star~\cite{Maund2015}. Similarly, the~progenitor of SN 2013ej also appears to be the near-superposition of a blue and a red star, although~their photocenters are just slightly displaced from one another~\cite{Fraser2014}.

\subsection{Progenitors of Low-Luminosity~SNe}\label{sec:lowlum}

Several of the SNe in Table~\ref{tab1} for which an RSG progenitor has been identified were low-luminosity events, including SN 2003gd~\cite{Hendry2005}, SN 2004A~\cite{Hendry2006}, SN 2005cs~\cite{Pastorello2009a}, SN 2008bk~\cite{VanDyk2012a,Lisakov2017}, SN 2018aoq~\cite{Tsvetkov2021}, and~SN 2022acko~\cite{Bostroem2023} (likely also SN 2006my~\cite{Maguire2010}). Interestingly, the~SED for these progenitors can be fit essentially with a bare photosphere, with~effective temperatures in the range of $T_{\rm eff} \sim 3400$--3700~K. This is typified by, e.g.,~the case of the SN 2018aoq progenitor (see Figure~\ref{sn18aoq}), for which the multi-band SED was fit by a 3500 K stellar atmosphere model with minimal, if~any, additional CSM extinction~\cite{ONeill2019}. 

\begin{figure}[H]
\includegraphics[width=10.5 cm]{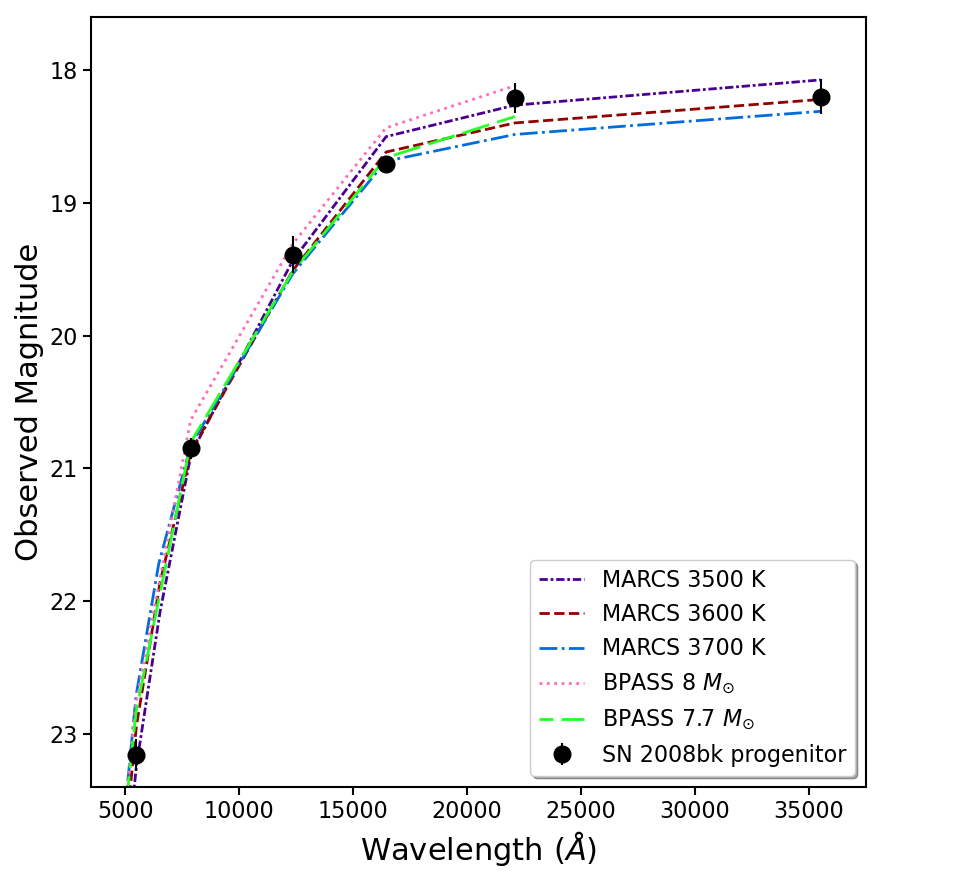}
\caption{Previously unpublished, corrected SED for the progenitor of SN 2008bk~\cite{VanDyk2012a} (see also~\cite{ONeill2021}). The~ground-based photometry for the star was adjusted for the contributions of fainter stars, visible with the {\sl HST}, within~the point-spread function in each band. Shown for comparison are {\tt MARCS} model stellar atmospheres at subsolar metallicity ($Z=0.010$; \cite{Gustafsson2008}) at 3500, 3600, and 3700 K. Also shown are the endpoint SEDs of BPASS model single-star evolutionary tracks at a similar metallicity, at~$M_{\rm ini}=7.7$ and $8\ M_{\odot}$. All of the models were reddened, assuming $A_V=0.054$ mag (from the Galactic foreground~\cite{Schlafly2011}, via the NASA/IPAC Extragalactic Database [NED]) and $R_V=3.1$.\label{sn08bk_sed} {The assumed distance to the SN is $3.44 \pm 0.12$ Mpc~\cite{VanDyk2012a}.}}
\end{figure}

We also illustrate this for the SN 2008bk progenitor in Figure~\ref{sn08bk_sed}. Correcting the photometry~\cite{VanDyk2012a} for the contributions of fainter stars, as~seen with the {\sl HST}, within~the ground-based point-spread function (PSF) to the overall brightness of the progenitor in each band (see also~\cite{ONeill2021}), we present the SED for the progenitor in the figure. For~comparison with the corrected SED, we show {\tt MARCS} stellar atmospheres at subsolar metallicity ($Z=0.010$ \cite{Gustafsson2008}) at $T_{\rm eff}=3600$, 3600, and~3700 K, as~well as the SEDs at the endpoints of two {\tt BPASS} model single-star evolutionary tracks~\cite{Stanway2018}. One can see that all of the photospheres provide excellent fits to the full observed SED. In~addition, the~endpoints of the 7.7 and $8\ M_{\odot}$ tracks are roughly consistent with the SED to {2.2} {\textmu}m. The~$T_{\rm eff}$ and $\log(L_{\rm bol}/L_{\odot})$ for the two model endpoints are 3555 K and 4.51 and 3550 K and 4.59, respectively. The~main point here is that any effect of extinction as a result of CSM dust on the stellar SED is minimal, at~least to wavelengths $\lesssim$ {3.5}\ \textmu m. The~luminosity implied for the SN 2008bk progenitor is consistent with the overall limiting luminosity for low-luminosity SNe II-P, i.e.,~$\log(L_{\rm bol}/L_{\odot}) \lesssim 4.7$. Fits to the SEDs of these progenitors generally---and~specifically in the cases of SN 2008bk and SN 2018aoq---appear to imply that $M_{\rm ini}\sim 8$--$10\ M_{\odot}$.

One notable and unresolved case is the low-luminosity SN 2018zd. This event was a collapse either as a result of an electron-capture (EC) reaction in a degenerate ONeMg core of an $M_{\rm ini} \sim 7$--$9.5\ M_{\odot}$ super-asymptotic giant branch (SAGB) star~\cite{Hiramatsu2021b} or~of a more standard Fe core in an $M_{\rm ini} \sim 12\ M_{\odot}$ RSG~\cite{Zhang2020}. The~ambiguity arises from a large uncertainty in the host distance ($\sim$6--18 Mpc). Assuming a distance on the short side,\mbox{ $\le$$10$ Mpc}, the~detected star at the SN position, indeed, had properties more consistent with a those of an SAGB than an RSG~\cite{Hiramatsu2021b}.

We point out that the SN 2022acko progenitor may have been straddling the SAGB/RSG boundary as well, with~$\log(L_{\rm bol}/L_{\odot}) \sim 4.3$--4.5~\cite{VanDyk2023b}. Although~SN 2005cs was only somewhat underluminous on the light-curve plateau, it suffered a large fall from the plateau to an extraordinarily underluminous radioactive tail, likely resulting from a very low $^{56}$Ni mass ($\sim$$3 \times 10^{-3}\ M_{\odot}$) synthesized in the explosion~\cite{Pastorello2009a}. The~luminosity estimates for the progenitor, $\log(L_{\rm bol}/L_{\odot}) \sim 4$--4.5~\cite{Maund2005a,Li2006}, are also indicative of a similarly lower-mass~progenitor.

\begin{figure}[H]
\includegraphics[width=10.5 cm]{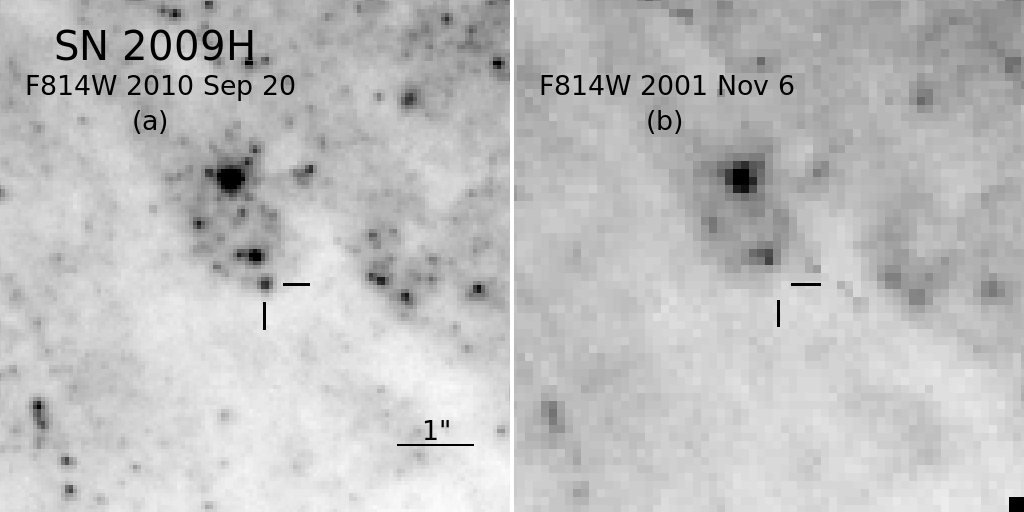}
\caption{(\textbf{a}) A portion of an {\sl HST\/} image of SN 2009H obtained in the F814W band on 2010 September 20 UT; the SN is indicated by tick marks. (\textbf{b}) A portion of an archival, pre-explosion {\sl HST\/} image, also in the F814W band, from 2001 November 6; the precise location of the SN site, based on astrometric registration with the image in panel (\textbf{a}) is indicated by tick marks. Both images are shown at the same scale and orientation. North is up, and east is to the~left.\label{sn2009h}}
\end{figure}

\begin{figure}[H]
\includegraphics[width=10.5 cm]{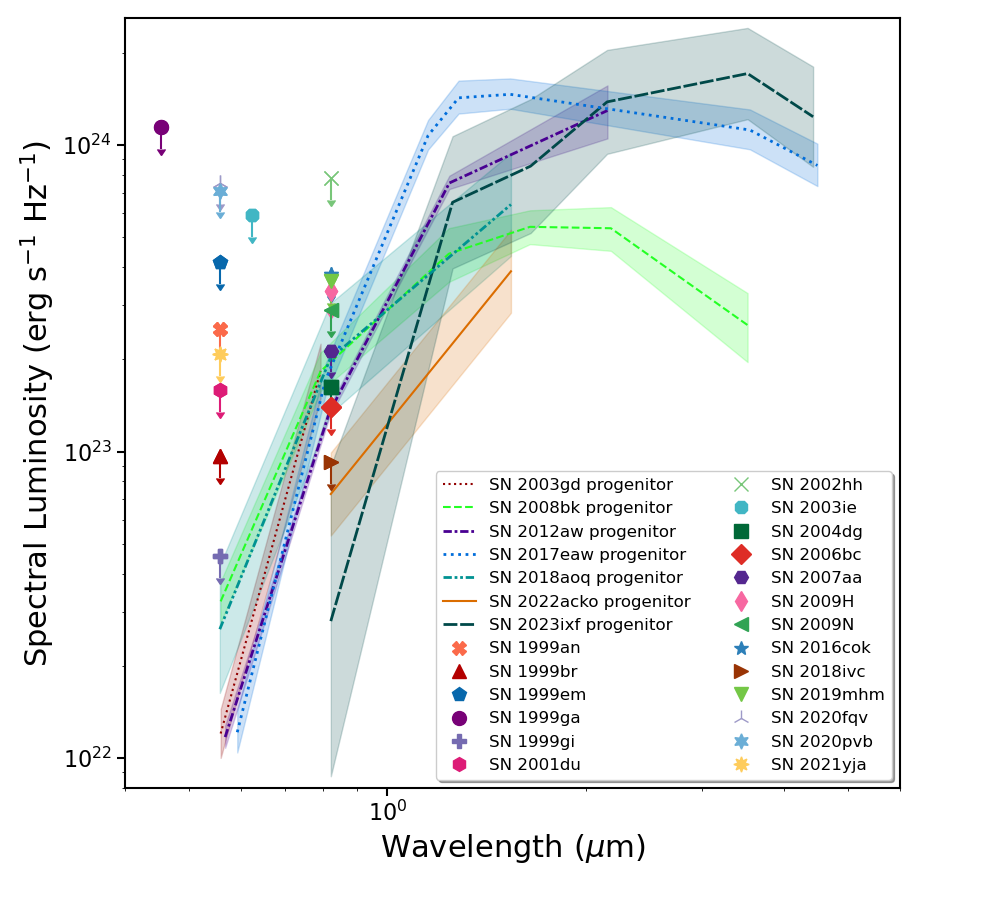}
\caption{Upper limits on RSG progenitor detections from Table~\ref{tab2}, shown relative to the intrinsic SEDs of the identified RSG progenitors shown in Figure~\ref{sedcomp}.\label{upperlimits}}
\end{figure}

What, if~anything, can we glean from the upper limits to progenitor detections presented in Table~\ref{tab2}? These limits were all estimated in optical photometric {\sl HST\/} bands, e.g.,~F555W or F814W. A~previously unpublished example for SN 2009H is shown in Figure~\ref{sn2009h}. What has been generally done in the literature is to infer a luminosity constraint by, e.g.,~adjusting the brightness limits in these bands with RSG bolometric \mbox{corrections~\cite{Levesque2005,Levesque2006,Davies2018}}, in~a similar way as for a number of the actual detections (e.g., \cite{Smartt2001}) (see Section~\ref{sec:direct}). However, this method may have led to luminosity underestimates~\cite{Healy2024,Beasor2025}. We show the upper limits from Table~\ref{tab2} in Figure~\ref{upperlimits} and compare these to the intrinsic SEDs of several of the direct detections (Figure~\ref{sedcomp}), with~the intent of imposing luminosity constraints from this~comparison. 
 
\unskip

What is readily obvious from Figure~\ref{upperlimits} is that the upper limits for several of the SNe do not provide meaningful constraints at all on the progenitor luminosities (contrary to what has been inferred in the literature for these events). However, some of the limits do, in~a limited way. Based on the blue end of the SEDs of the progenitor detections and the luminosities associated with those stars (see Table~\ref{tab1}), we can infer at least relatively useful, although~approximate, luminosity upper limits, and~for several cases, we can also place approximate lower limits. We provide these estimates in Table~\ref{tab2}. The~lowest lower-luminosity constraints arise from the observed limits being indicative of a progenitor at least as luminous as the detected low-luminosity progenitors (e.g., SN 2008bk; $\log[L_{\rm bol}/L_{\odot}] \gtrsim 4.5$). The~upper luminosity constraints arise from comparison with, e.g.,~ the SN 2023ixf progenitor SED; therefore, the progenitor was likely no more luminous than that ($\log[L_{\rm bol}/L_{\odot}] \lesssim 5.1$; see Section~\ref{sec:sn2023ixf}). The~optically based limits have little constraining power (since most of the emission from an RSG is at longer wavelengths) and, had detection limits been available, particularly in the near-IR range, we would likely have been able to place far more stringent luminosity constraints, again, treating the known progenitor SEDs as~templates.

\begin{figure}[H]
\includegraphics[width=10.5 cm]{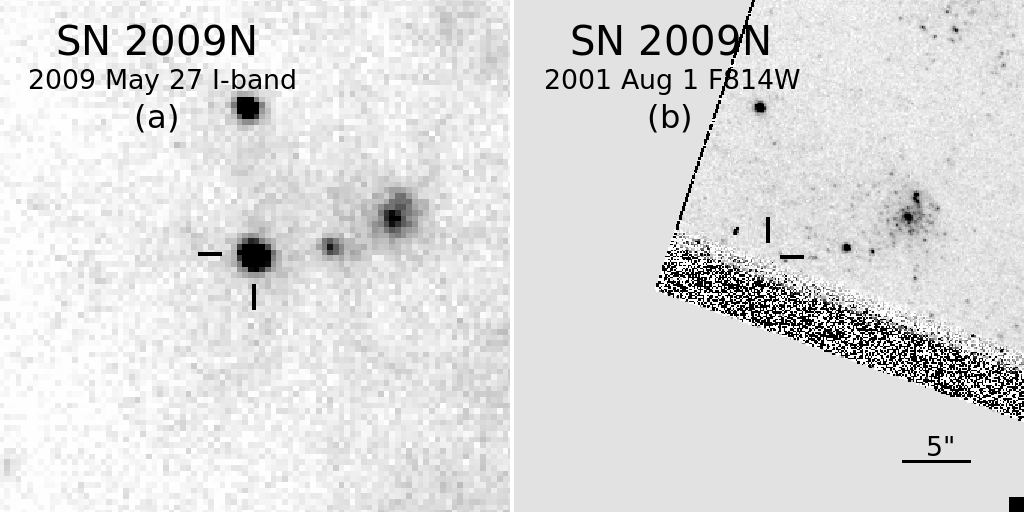}
\caption{(\textbf{a}) A portion of an image of SN 2009N obtained with the Palomar Observatory 1.5 m telescope (in California, USA) in the $I$ band on {27  May 2009;} the SN is indicated by tick marks. (\textbf{b})~A portion of an archival, pre-explosion {\sl HST\/} image in the F814W band from {1 August 2001;} the approximate location of the SN site, based on astrometric registration with the image in the panel (\textbf{a}), is indicated by tick marks. Note the extreme proximity to the detector edge of the site of the SN progenitor. These images were not previously presented~\cite{Takats2014}. Both images are shown approximately at the same scale and orientation. North is up, and east is to the~left.\label{sn2009n}}
\end{figure}

With regards to the low-luminosity SNe II-P, the~limits on detection of a progenitor only likely pertain to SN 2009N (see Figure~\ref{sn2009n}), which had spectral properties similar to low-luminosity events, yet had photometric similarities to more intermediate-luminosity SNe II-P~\cite{Takats2014}.

The nearby ($\sim$14 Mpc) SN 2020jfo appears to be something of an unusual puzzle. The light-curve properties of the SN are consistent with more luminous SNe, including a short plateau duration of $\lesssim$$65$ days~\cite{Sollerman2021,Teja2022,Ailawadhi2023,Kilpatrick2023a}; however, a~single-band progenitor detection at {\sl HST\/} F814W was underluminous relative to what would be expected for luminous, relatively massive RSGs, which could also imply enhanced CSM dust~\cite{Sollerman2021}. Furthermore, analyses of the late-time spectra were indicative of a star with a moderate initial mass of $\sim$$12\ M_{\odot}$ \cite{Sollerman2021,Teja2022,Ailawadhi2023,Kilpatrick2023a}. However, the~progenitor may also have been detected with the {\sl Spitzer\/} telescope, and, as~measured at these wavelengths, the~total SED could be fit with a quite cool ($\sim$2900 K), low-luminosity ($\log[L_{\rm bol}/L_{\odot}] \sim 3.9$) blackbody; even including a consideration of CSM dust raised the luminosity estimate only somewhat to~$\log(L_{\rm bol}/L_{\odot}) \sim 4.1$, which further implies a low initial mass~\cite{Kilpatrick2023a}. Based on all of this, it is unclear whether SN 2020jfo should really be considered a low-luminosity SN, even though the inferred properties of its progenitor, at~this point, at least, are consistent with those of other low-luminosity SN progenitor stars (see also~\cite{Utrobin2024}).

In summary, all in all, low-luminosity SNe II-P appear to arise from relatively lower-luminosity RSGs, for~which dust in the CSM is likely relatively limited in quantity. One can infer that these RSGs are also at the low end of the possible range in initial mass for progenitors, i.e.,~$M_{\rm ini} \sim 8$--$30\ M_{\odot}$. Some of these progenitors may have been in a transitional state between EC-SN and Fe core-collapse~SN.


\subsection{Progenitors of More Luminous~SNe}\label{sec:highlum}

The remaining SNe in Table~\ref{tab1} with identified progenitors appear to have arisen from RSGs with higher luminosities ($\log(L_{\rm bol}/L_{\odot}) \gtrsim 4.7$) based on the analyses of the stars' observed properties {(although, this demarcation is somewhat arbitrary and not yet well-defined)}. Since the CSM dust production rate appears to be correlated with $L_{\rm bol}$ \cite{Massey2005,Verhoelst2009} (the mass of CSM dust may also increase as the RSG progenitor approaches explosion~\cite{Beasor2016}), the~effect of dust on the observed SED is expected to be larger for higher-luminosity~progenitors.

If one examines Figure~\ref{sedcomp}, then this is effectively what one sees. The~presence of and necessity for 
including CSM dust became particularly apparent when attempting to fit the observed SED of the SN 2012aw progenitor~\cite{Kochanek2012}. The reddened overall shape of the SED, compared to that of a bare photosphere, is evident, as~the dust in the CSM reprocesses the UV/blue photospheric emission into the infrared range. This was even more conspicuous in the detailed multiband SED of the SN 2017eaw progenitor~\cite{Kilpatrick2018,VanDyk2019}. The~lone attempt to characterize the SN 2024ggi progenitor so far resulted in a \mbox{luminosity $\log(L_{\rm bol}/L_{\odot}) \sim 4.9$ \cite{Xiang2024b}.} All of these RSG progenitors had luminosities at explosion approaching 100,000~Suns.

As one can see from Figure~\ref{sedcomp} the SED for the SN 2004et progenitor is unusual relative to the SEDs of the other SN progenitors (including for low-luminosity SNe). It looks quite unlike the expected SED of an RSG --- the luminosity of the star in the blue is far higher than that of any of the other progenitors presumed to be RSGs. The~SN 2004et progenitor was characterized solely based on ground-based image data. Subsequent observations of the field with the {\sl HST\/} revealed that other, fainter stars within the PSF may have contaminated the light from the observed progenitor itself. However, even accounting for this contamination only reduces the observed flux in each band by $\sim$$0.3$ mag or so, and~the SED is still better characterized by a luminous star hotter than the M type, i.e.,~a yellow supergiant~\cite{Crockett2011}. What may be the case is that the detected star is actually a composite of more than one star. We will not really know until the SN has significantly faded; however, this may well be a long wait, since SN 2004et continues to exhibit signs of CSM interaction, as~well as the formation of dust~\cite{Shahbandeh2023}.

The progenitor requiring the largest amount of CSM dust to date in its characterization is SN 2023ixf, and~we reserve a separate discussion for the next subsection (Section~\ref{sec:sn2023ixf}).

Have there been SN II-P progenitors more luminous than $\log(L_{\rm bol}/L_{\odot})\approx 5.0$? The modeling of the two-band detection of the SN 2009ib progenitor in {\sl HST\/} archival data led to an estimate of $\log(L_{\rm bol}/L_{\odot}) = 5.12 \pm 0.14$ \cite{Takats2015}, although~the SN itself was assessed possibly to be of low luminosity (or, at~least, at~the transition from low to more normal luminosity). Based on a single {\sl HST\/} band (F814W) detection, however, the~progenitor of SN 2012ec was estimated to have had $\log(L_{\rm bol}/L_{\odot}) = 5.15 \pm 0.19$ \cite{Maund2013}, which would imply that it was one of the most luminous (and potentially most massive) known RSG progenitors, which, unlike for SN 2009ib, is not entirely inconsistent with the observed properties of the SN itself~\cite{Maund2013,Barbarino2015}.

Whereas the overwhelming majority of the identified progenitors were well-isolated spatially, we have already noted that the SN 2008cn progenitor candidate~\cite{EliasRosa2009} was likely a blend of the RSG with a superposed (or companion) blue star~\cite{Maund2015} (we note that the progenitor candidate was distinctly variable with a period $\sim$64 days~\cite{EliasRosa2009}) and~that the SN 2013ej progenitor was partially confused with the presence of a blue neighbor~\cite{Fraser2014}. Therefore, it was very difficult to measure and accurately characterize the brightness of the star before explosion, despite indications that both were quite luminous. Although~data were scant, the~observed properties of SN 2008cn indicated that it was a luminous SN II-P~\cite{EliasRosa2009}. SN 2013ej was somewhat unusual in~that it was a slow-rising~\cite{Valenti2014} and fast-declining, luminous SN II-P or II-L~\cite{Bose2015,Haung2015,Dhungana2016,Yuan2016} that was also powered, at later times, by the interaction of the SN shock with the CSM~\cite{Mauerhan2017}. The~observed properties of SN 2009hd, particularly its fast decline, pointed to it being a possible SN II-L; substantial uncertainty existed as to whether the progenitor was actually identified due to the level of confusion in its environment, even at {\sl HST\/} resolution, and~the fact the SN observations indicated that the line of sight to the star is heavily dust-obscured~\cite{EliasRosa2011}. Nevertheless, the~likelihood is that it was a luminous~RSG.

As discussed in Section~\ref{sec:lowlum}, approximate constraints on the luminosity of a progenitor can be estimated from the observed limits on the progenitor's detection. Lower luminosity constraints can be placed based on comparison of the detection limits with the least dusty SEDs of the lower-luminosity progenitors, and~upper luminosity constraints can be established based on comparison with the most dusty SED, i.e.,~for SN 2023ixf. In~three cases of moderate- to higher-luminosity SNe II (SN 2004dg~\cite{Harutyunyan2008}, SN 2007aa~\cite{Anderson2014,Gutierrez2017}, and SN 2018ivc~\cite{Bostroem2020}), we can infer that the star was at least as luminous as the SN 2008bk progenitor ($\log[L_{\rm bol}/L_{\odot}] \gtrsim 4.5$) and, in one case, (SN 2006bc), at~least as luminous as the SN 2012aw progenitor ($\log[L_{\rm bol}/L_{\odot}] \gtrsim 4.8$). 


\subsection{The SN 2023ixf Progenitor as a Special~Example}\label{sec:sn2023ixf}

SN 2023ixf occurred in the nearly face-on Messier 101 (NGC 5457) at a distance of 6.85~Mpc, in~close proximity within the host to giant H~{\sc ii} region NGC 5471. As~such, the~SN generated tremendous excitement very soon after discovery, at~14.9 mag, with~classification as an SN II, and~immediately became the target of intense scrutiny by a number of investigators working across the full gamut of wavelengths. Given the closeness and prominence of the host and interest in the nearby {H~{\sc ii}} region, a~veritable treasure trove of archival data, both ground- and space-based, were available.  In particular, the host had been observed over numerous epochs with {\sl Spitzer}. (In fact, due to the relative scarcity of pre-explosion optical imaging data in which the star was detected, {\sl HST\/} contributed comparatively little to our knowledge of this SN progenitor.) It became readily evident that the star (which could be straightforwardly isolated in {\sl Spitzer\/} data with a high degree of certainty based on its equatorial coordinates alone) was highly variable. In~fact, it was possible to describe the star classification as a semi-regular, LPV (period $P \approx 1000$--1100 days~\cite{Jencson2023,Soraisam2023}), not unlike many known RSGs in the local group. This, in~itself, was an astonishing~discovery. 

The existence of exquisite multi-epoch photometry at various bands over years before explosion also helped rule out any dramatic changes in the star's luminosity up to days the SN, effectively ruling out the occurrences of any pre-SN outbursts, eruptions, or~other catastrophic late-time mass loss~\cite{Jencson2023,Soraisam2023,Flinner2023,Hiramatsu2023,Dong2023,Ransome2024}. Therefore, the~origin of the dense CSM inferred around SN 2023ixf must have occurred via some other~mechanism.


With the wealth of pre-explosion data, the proximity and brightness of the SN, and~the relative isolation and ease of identification of the progenitor candidate, a~number of investigators undertook the construction of appraisals of the progenitor's properties. Beyond~the general agreement that resulted in the analysis of the star's variability, as~mentioned above, and~the overall recognition that the CSM was especially dusty, the~approaches of these various investigations and their results were, by and large, disparate for~a number of reasons. From~the start, the~measurement techniques for obtaining photometry from the existing {\sl HST}, {\sl Spitzer}, and~ground-based near-IR data significantly diverged: For the {\sl HST\/} data, most studies employed the {\tt Dolphot\/} package {v2.0}~\cite{Dolphin2016}, while one~\cite{Pledger2023} used {\tt DAOPHOT\/}~\cite{Stetson1987} and yet another used a tool developed by the authors~\cite{Qin2024}. In principle, everyone should have obtained the same numbers; however, even those using {\tt Dolphot\/} arrived at different values based on different parameter settings, pre-processing steps, and~so on. For~measurements from the {\sl Spitzer\/} image data, one study applied aperture photometry to the pipeline-processed mosaics~\cite{Xiang2024a}; another applied {\tt DAOPHOT\/} PSF fitting to these mosaics~\cite{Jencson2023}, another employed {\tt DoPHOT\/} \cite{Schechter1993} on the mosaics~\cite{Niu2023}, while yet another~\cite{Soraisam2023} used the {\tt MOPEX\/} \cite{Makovoz2005b} and {\tt APEX\/} MultiFrame~\cite{Makovoz2005a} packages {v18.5.6} as~advised by the {\sl Spitzer\/} project. Needless to say, again, a~range of output values was obtained. In~\mbox{Figure~\ref{sedcomp}}, we show the SED constructed by~\citet{VanDyk2024} from {\sl HST\/} F814W to {\sl Spitzer\/} 4.5 {\textmu}m, for which the uncertainties included the estimated amount of variability in each~band.

Next, the~approaches to model fitting to the various SEDs differed. Several of the studies~\cite{Kilpatrick2023b,Niu2023,Neustadt2024,Xiang2024a,Ransome2024,Qin2024} used the dust radiative-transfer code ({\tt DUSTY}) \cite{Ivezic1999}, together with {\tt MARCS} atmospheres~\cite{Gustafsson2008} as the input light source; another~\cite{VanDyk2024} used {\tt PHOENIX} atmospheres~\cite{Kucinskas2005} instead. Two studies~\cite{Jencson2023,VanDyk2024} used {\tt GRAMS} models~\cite{Sargent2011,Srinivasan2011}, in~lieu of or in addition to {\tt DUSTY}. About half of the studies adopted silicate-rich dust for the CSM, whereas roughly the other half adopted carbon-rich dust (although the CSM around known RSGs tends to have little to no C-rich dust~\cite{Verhoelst2009}).

\begin{figure}[H]
\includegraphics[width=10.5 cm]{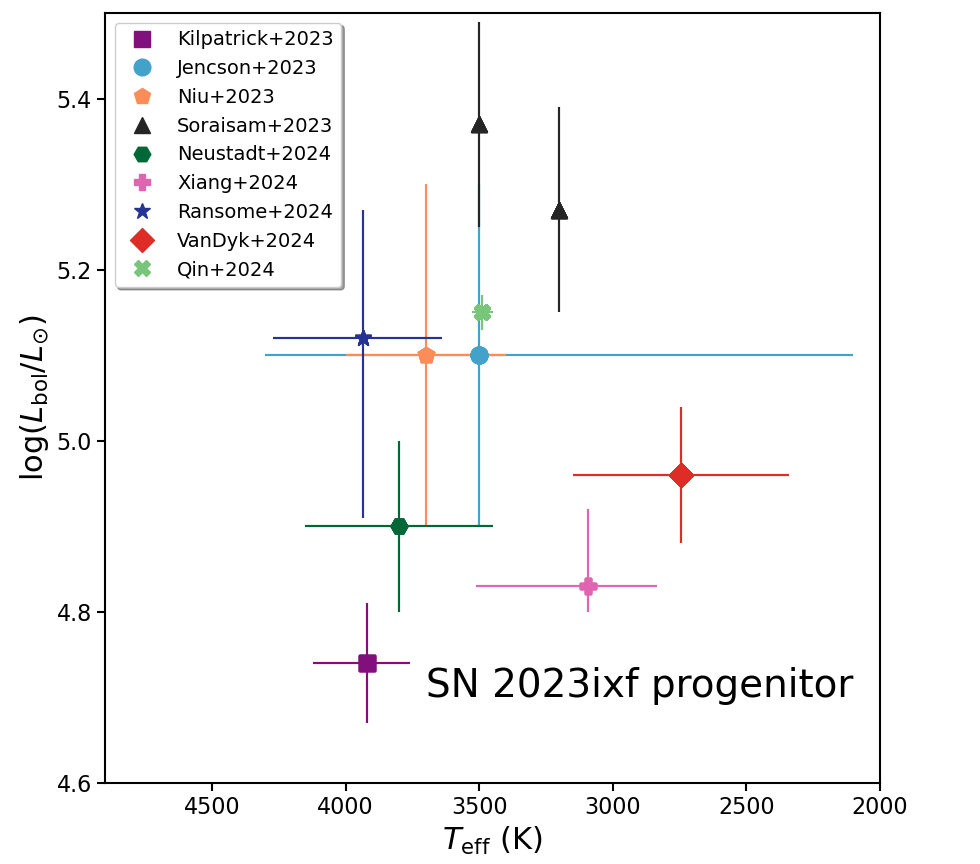}
\caption{Comparison of the various estimates of the bolometric luminosity ($\log(L_{\rm bol}/L_{\odot})$) and~the effective temperature ($T_{\rm eff}$) of~the SN 2023ixf RSG progenitor based on either SED fitting or via a P-L relation applied to the variable star~\cite{Kilpatrick2023b,Jencson2023,Niu2023,Soraisam2023,Neustadt2024,Xiang2024a,Ransome2024,VanDyk2024,Qin2024}.\label{sn23ixf_lum}}
\end{figure}

The net result of all of these various assumptions and approaches is an astonishing level of disagreement on the bolometric luminosity and effective temperature of the star years to days before it exploded (see Figure~\ref{sn23ixf_lum}). This might have been expected if little data had been available to construct the star's SED. However, in~this case, we had an astounding level of multi-band data available, from~the UV to the mid-infrared band. Given that, we should have been able to sharply define this one star's characteristics; therefore, it is frustrating that the results have not converged. Technically, from~a statistical standpoint, the~various values for $T_{\rm eff}$ and $\log(L_{\rm bol}/L_{\odot})$ do agree with each other in the ensemble, to~within all of their estimated uncertainties; one could compute a weighted mean and uncertainty in each of $3478 \pm 392$~K and $5.09 \pm 0.17$. However, it is disappointing that, given the very high quality of the available data for the star, we have to place such large overall uncertainties on these inferred~values.

How we~\cite{Soraisam2023} approached this analysis was to carefully measure the fluxes or~estimate upper limits of detection based on the available {\sl HST}, {\sl Spitzer}, and~ground-based near-IR imaging data over all of the available epochs, then perform a variability analysis of the resulting light curves. We estimated a period of $1091 \pm 71$ days for the star over years before explosion. From~this period, assuming two different $T_{\rm eff}$ values, we inferred bolometric luminosities from an established RSG period--luminosity (P-L) relation, with both values being on the high side ($\log[L_{\rm bol}/L_{\odot}] \sim 5.3$--5.4). Next, we averaged the various epochs in the IR to arrive at global values, including the amplitude of the flux variability into the overall uncertainty in~each band (for the single {\sl HST\/} band in which the star was detected, we adopted a representative amplitude based on other cool variable giants). Then, we formally fit the SED with both {\tt GRAMS} and {\tt DUSTY}, assuming Si-rich dust, and~obtained consistent results (and likely more stringent ones than from the P-L relation) of $\log(L_{\rm bol}/L_{\odot})$ = \mbox{4.88--5.04} and $T_{\rm eff}=2340$--3150 K (68\% credible intervals) \cite{VanDyk2024}. As~can be seen in Figure~\ref{sn23ixf_lum}, these are on the lower-luminosity and (considerably) cooler side. Lest it be a consideration that our estimated properties were fanciful and unrealistic, we found a direct galactic analog, RSG IRC $-${10414}, with~strikingly similar properties~\cite{Gvaramadze2014,Messineo2019}. (We did not necessarily intend, in our analysis, to suggest that IRC $-${10414} is also nearing explosion.) Interestingly, we concluded that the SN 2023ixf progenitor, like IRC $-${10414} (and Betelgeuse~\cite{Noriega1997} and $\mu$ Cep~\cite{Cox2012}), may have had an interstellar bow~shock.

Even with the surprising disparity in the inferred properties of the SN 2023ixf progenitor (with the overall disagreement, it is unfortunate that the community will likely just randomly pick from the various results when referencing this progenitor in the future), we were able to characterize this candidate star in extraordinary and unprecedented detail. We will likely not have the opportunity for a similar or~superior breakthrough until an SN occurs in a host galaxy with multi-band {\sl JWST\/} coverage. Although~the level of certainty is exceedingly high that the candidate was, indeed, the progenitor, this still requires confirmation when the SN has sufficiently faded (see the discussion in Section~\ref{sec:intro}). 


\section{Discussion}

The key observationally driven advances in our understanding of SN II-P progenitors and~their link to RSGs have only been made possible by the direct identification of progenitors in pre-explosion archival images. The~statistics are still amazingly too small to draw any broad conclusions about SN II-P progenitors, despite previous attempts at this. Accumulating sufficient statistics on SN progenitors requires both time and patience; thus, every new example is highly welcomed and informative. Although~upper limits of detection have provided some coarse indications of progenitor characteristics, only actual progenitors and progenitor system identifications will continue to advance this line of study in finer~detail.

{As a community, we have generally found that low-luminosity SNe arise from lower-luminosity ($\log[L_{\rm bol}/L_{\odot}] \sim 4$--4.7; and lower $M_{\rm ini}$) RSGs~\cite{Lisakov2018}. Given a normal initial mass function within host galaxies in the local Universe, we would expect these progenitors to be predominant. These RSGs appear to have less (and~less dusty) CSM prior to explosion, and~in a few cases, the~stars may dwell near the boundary of RSG and SAGB, perishing as either a core-collapse or electron-capture event. Whereas low-luminosity SNe II-P possess a certain degree of observational homogeneity, both in photometric and spectroscopic properties~\cite{Pastorello2004,Spiro2014}, higher-luminosity SNe are more heterogeneous in their properties, preventing us from clearly distinguishing what is ``normal'' from what is more extreme. If anything, a~continuum of properties likely exists, starting at the low-luminosity \mbox{floor~\cite{Anderson2014,Faran2014,Pejcha2015,Valenti2016,Gutierrez2017}}. We likely see this reflected in the diversity of observed progenitor characteristics for RSGs with $\log(L_{\rm bol}/L_{\odot}) \gtrsim 4.7$. Suffice it to say that the~mass, extent, and~dust content of CSM is substantially higher for progenitors of more luminous explosions.}

Target-of-opportunity programs with both the {\sl HST\/} and AO, observing young SNe II-P to identify progenitor candidates, have been ongoing for the last two decades and necessitate the virtual equivalent of a long-term program because~of the relatively low rate of viable candidates. ({\sl JWST\/} has yet to be purposely used in this capacity, although it has serendipitously it has~\cite{VanDyk2023b}.) Over time we, as~a community, have successfully amassed constraints on dozens of interesting SNe and their progenitors. Every new progenitor identification is precious, and~each has greatly advanced our understanding of SN explosion physics, stellar evolution, and~progenitor mass-loss history, particularly by putting the characteristics of each SN in context with the properties of its directly identified~progenitor.


\subsection*{The ``Red Supergiant~Problem''}\label{sec:rsg_problem}

\begin{figure}[H]
\includegraphics[width=10.5 cm]{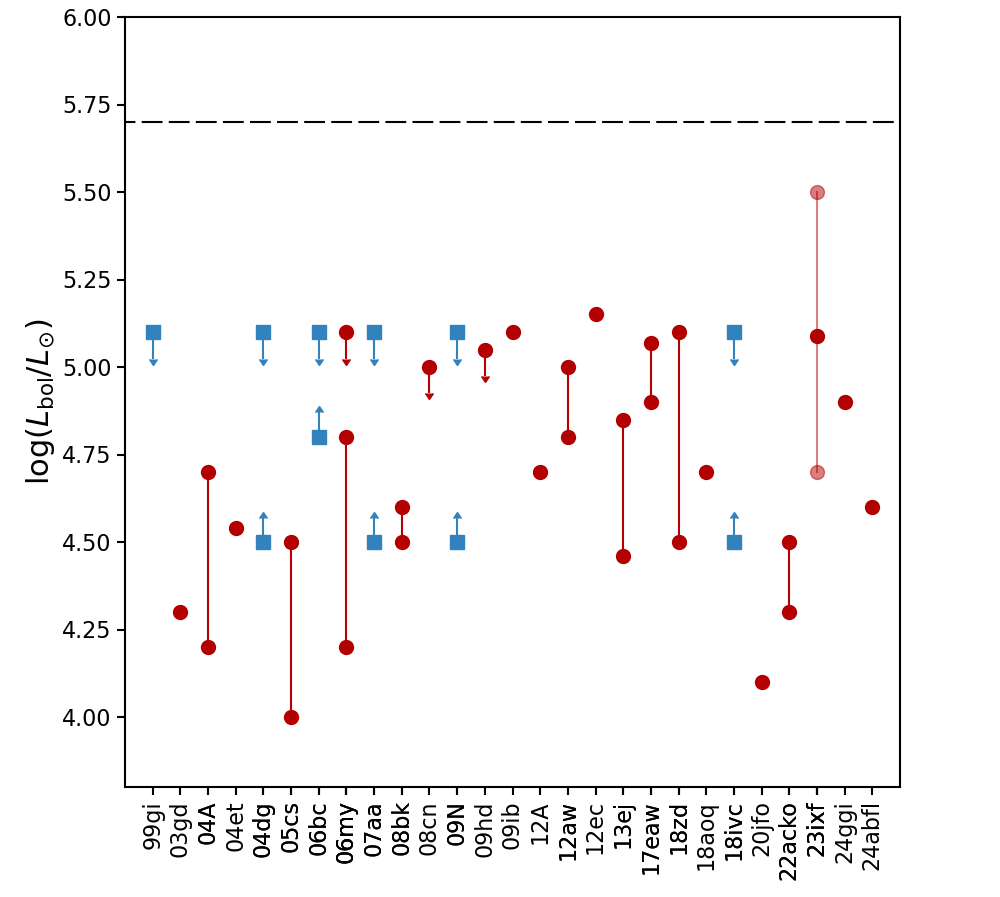}
\caption{Comparison of the estimates of the bolometric luminosity ($\log(L_{\rm bol}/L_{\odot})$) for~all of the SNe II-P considered here. The~red circles are the estimates from direct progenitor identification, whereas the blue squares are estimates we made here based on the non-detection limits. In~the case of SN 2023ixf, we show the range in the luminosity estimates, as~well as the weighted mean (see Section~\ref{sec:sn2023ixf}). Also shown is the Humphreys--Davidson limit~\cite{Humphreys1979} (see also Humphreys and Jones \& Humphreys in this volume).\label{prog_lum_comp}}
\end{figure}

This brings us to one last topic on RSGs as SN progenitors. As~one can see from Figure~\ref{prog_lum_comp}, the luminosities derived for the SN progenitors, both from direct identifications and from limits on detection, all tend to top out at $\log(L_{\rm bol}/L_{\odot}) \lesssim 5.2$ (with the possible exception of the upper range of the luminosity estimate for the SN 2023ixf progenitor) and~generally fall well below the empirical Humphreys--Davidson limit~\cite{Humphreys1979} on the luminosities of local-group cool supergiants ($\log(L_{\rm bol}/L_{\odot}) \sim 5.7$; see also Humphreys and Jones \& Humphreys in this volume). The~apparent fact that no SN II-P progenitor has, so far, been found closer to this limit, even though such luminous RSGs exist in the local Universe, is a puzzle. Why are we not finding highly luminous RSG SN progenitors? This so-called ``red supergiant problem'' has been debated intensively in the literature~\cite{Smartt2009a,Smartt2015,Davies2020a,Kochanek2020,Davies2020b}, and~various explanations have been offered to help solve this conundrum: from CSM dust~\cite{Walmswell2012} to~bolometric correction~\cite{Davies2018}, statistics~\cite{Davies2020b,Strotjohann2024}, and alternative evolutionary pathways for the highest-luminosity (most initially massive) RSGs~\cite{Sukhbold2014}. The fact that the fate of the most massive RSGs may depend on the core compactness parameter~\cite{Horiuchi2014} can lead to mass ranges that terminate in failed explosions, leading directly to black-hole formation~\cite{OConnor2011,Kochanek2015}.

The search was then underway for RSGs that simply failed to explode~\cite{Gerke2015,Adams2017b} or experienced very low-energy SNe~\cite{Lovegrove2013}. A~promising candidate source, ``NGC 6946-BH1'', was isolated~\cite{Adams2017a}, although~according to analysis of recent {\sl JWST\/} observations, its origin remains ambiguous~\cite{Kochanek2024,Beasor2024a}{; possibilities include a stellar merger~\cite{Beasor2024a} manifested as an intermediate-luminosity optical transient~\cite{Kashi2017} or luminous red nova~\cite{Steinmetz2025}.}

Possibly a simpler means of explaining, or attempting to explain away, the RSG problem is by taking into account the full SEDs of RSGs, from the optical to the mid-IR. Since many of the SN progenitor identifications are in a single band, the question asked is, how much divergence is the inferred luminosity from the actual luminosity; it has essentially been argued that the inference is underluminous, and~taking this into account, one could statistically eliminate the existence of an ``RSG problem~\cite{Healy2024,Beasor2025}'', although the assumed luminosities for identified progenitors differ to a fair extent from those we have presented~here.

We would argue that the RSG problem remains to be adequately challenged observationally. The sample still needs to be at least doubled~\cite{Davies2020a}. Therefore, every new SN II with a progenitor identification further augments the luminosity (mass) distribution.

\unskip 

\vspace{6pt} 

\funding{This research received no external~funding.}

\dataavailability{All data used in this review are publicly available.} 

\conflictsofinterest{The author declares no conflicts of~interest.} 

\abbreviations{Abbreviations}{
The following abbreviations are used in this manuscript:\\

\noindent 
\begin{tabular}{@{}ll}
3D & Three-Dimensional\\
CSM & Circumstellar Matter (or Medium)\\
EC & Electron Capture\\
HST & Hubble Space Telescope\\
IR & Infrared\\
JWST & James Webb Space Telescope\\
II-L & II-Linear\\
II-P & II-Plateau\\
LPV & Long-Period Variable\\
NED & NASA/IPAC Extragalactic Database\\
Ni & Nickel\\
P-L & Period-Luminosity\\
PSF & Point Spread Function\\
RSG & Red Supergiant\\
SAGB & Super Asymptotic Giant Branch\\
SED & Spectral Energy Distribution\\
SN & Supernova\\
SNe & Supernovae
\end{tabular}
}

\begin{adjustwidth}{-\extralength}{0cm}

\reftitle{References}

\PublishersNote{}
\end{adjustwidth}
\end{document}